# Engineering Phonon-Qubit Interactions using Phononic Crystals

Kazuhiro Kuruma[1, a)], Benjamin Pingault[1,2], Cleaven Chia[1], Michael Haas[1], Graham D Joe[1], Daniel Rimoli Assumpcao[1], Sophie Weiyi Ding[1], Chang Jin[1], C. J. Xin[1], Matthew Yeh[1], Neil Sinclair[1], and Marko Lončar[1, b)]

[1] *John A. Paulson School of Engineering and Applied Sciences, Harvard University, Cambridge, Massachusetts 02138, USA*
[2] *QuTech, Delft University of Technology, PO Box 5046, 2600 GA Delft, The Netherlands, EU*

a) E-mail: kkuruma@seas.harvard.edu
b) E-mail: loncar@seas.harvard.edu

**Abstract**
**The ability to control phonons in solids is key for diverse quantum applications, ranging from quantum information processing to sensing. Often, phonons are sources of noise and decoherence, since they can interact with a variety of solid-state quantum systems. To mitigate this, quantum systems typically operate at milli-Kelvin temperatures to reduce the number of thermal phonons. Here we demonstrate an alternative approach that relies on engineering phononic density of states, drawing inspiration from photonic bandgap structures that have been used to control the spontaneous emission of quantum emitters. We design and fabricate diamond phononic crystals with a complete phononic bandgap spanning 50 - 70 gigahertz, tailored to suppress interactions of a single silicon-vacancy color center with resonant phonons of the thermal bath. At 4 Kelvin, we demonstrate a reduction of the phonon-induced orbital relaxation rate of the color center by a factor of 18 compared to bulk. Furthermore, we show that the phononic bandgap can efficiently suppress phonon-color center interactions up to 20 Kelvin. In addition to enabling operation of quantum memories at higher temperatures, the ability to engineer qubit-phonon interactions may enable new functionalities for quantum science and technology, where phonons are used as carriers of quantum information.**

Engineering the interactions of quantum systems with phonons/vibrations is an important task for a wide range of disciplines, including quantum information[1–3], optoelectronics[4], metrology[5], energy harvesting[6], and sensing[7,8]. Coherent phonons can play an important role as carriers of quantum information[9,10], whereas thermal phonons can negatively impact the coherence properties of many quantum systems even at single-phonon levels, and eventually limit the coherence of quantum devices[11,12]. The most common approach to address this issue is to operate at reduced temperatures, typically in the milli-Kelvin range, to decrease the population of thermal phonons[9,13,14]. However, this approach requires complex and expensive cryogenic systems and does not mitigate the effects of local heating due to driving fields used to interact with the quantum system[15,16]. Most importantly, this approach does not allow for precise control of the local interactions between phonons and quantum systems, including those required for implementing quantum operations and measurements. It is thus important to develop alternative approaches to control thermal phonons and their interactions with quantum systems.

An emerging approach relies on engineering the phononic local density of states (LDOS) by fabricating micro/nanostructures. In particular, phononic crystals (PnCs)[17] with periodic modulations of elastic properties have been intensively studied since they can possess phononic bandgaps, a frequency range in which phonons cannot propagate. This is analogous to photonic crystals that support photonic bandgaps for electromagnetic waves[18]. To date, demonstrated PnCs typically operate at relatively low frequencies (≲10 GHz) and have mainly been used to realize phononic shields[19] to reduce mechanical losses, as well as optomechanical cavities[20,21] and waveguides[22,23]. On the other hand, PnCs offer a great opportunity to tailor the interactions between thermal phonons and emerging qubits or quantum emitters, and could suppress decoherence processes induced by high frequency phonons[11,12,24], thus enabling operation at elevated temperatures. However, tailoring PnCs to specific qubits is challenging since it requires nanofabrication strategies for non-standard materials (in which qubits are hosted), which maintain or ideally improve qubits' properties (e.g. coherence, optical stability). Furthermore, a high degree of control over the frequencies of the targeted phonons is required, which poses additional challenges in particular when those frequencies are high (50-100GHz), thereby relying on features in the 10-100 nm range. As a result, previous attempts at control of the interaction between bulk/thermal phonons and a quantum system relied on tuning the energy levels of the quantum system[25,26] or lacked the capacity to tailor the LDOS[27].

Here, we demonstrate free-standing diamond PnCs with ~20nm features, which support a ~20 GHz-wide complete bandgap centered at ~60 GHz and use them to mitigate the phonon-induced relaxation of a single quantum system. Our quantum system is a negatively charged silicon-vacancy (SiV) color center in diamond. The SiV is a leading quantum memory candidate for the development of quantum networks[28,29] due to its excellent optical properties and an optically accessible spin. However, the SiV also features high sensitivity to strain and mechanical vibrations[26,30,31], and typically requires operation at milli-Kelvin temperatures to decouple it from the thermal bath[13]. Here, we use single SiV centers as an optical probe of the phononic LDOS, and show evidence of the high-frequency phononic bandgap. Our work draws inspiration from previous work on controlling photonic LDOS using photonic crystals to inhibit spontaneous light emission from quantum emitters[32,33]. By suppressing spontaneous single-phonon processes in SiV centers using PnCs, we experimentally show that orbital relaxation rates within the phononic bandgap can be more than one order of magnitude lower than in bulk. This large reduction in the relaxation rates using PnCs may allow operation of SiVs at elevated temperatures.

# Results
## Design and Fabrication of diamond PnCs

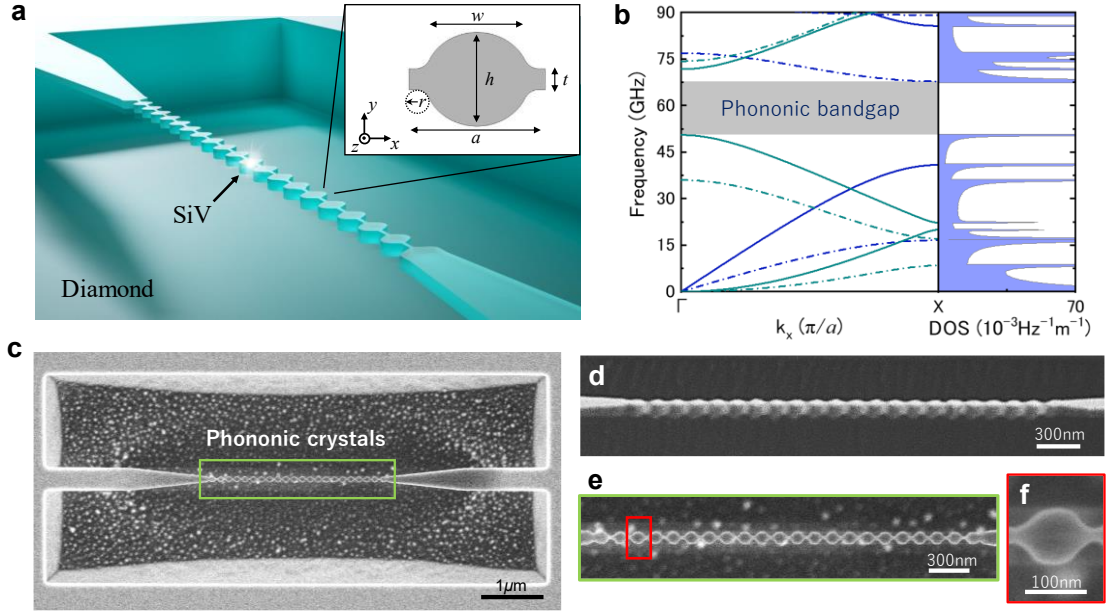

**Fig. 1 | PnCs in single-crystal diamond. a,** Illustration of free-standing PnCs supported with tapered waveguides at both ends. Inset shows the unit cell of the PnCs, consisting of an elliptical block and two small tethers with a width of $t$. The major and minor axes of the elliptical block and the lattice constant are $w$, $h$, and $a$, respectively. The small circle with a radius of $r$ is to account for the rounding of the connection between the block and the tethers due to fabrication imperfection. **b,** Left: Simulated phononic band diagram corresponding to the fabricated PnC. Unit cell parameters were obtained by scanning electron microscopy (SEM) of the fabricated structure (shown c-f). The solid (dashed) blue line indicates acoustic modes with even (odd) symmetry along y- and z-axes, while solid (dashed) green indicates modes with even (odd) symmetry along y-axis and odd (even) symmetry along z-axis. The gray region indicates a complete phononic bandgap with a center frequency of 59.1 GHz and width of 17.3 GHz. Right panel shows the corresponding phononic density of states (DOS). **c-f,** SEM images showing top (**c**) and tilted (**d**) views of fabricated suspended PnCs, consisting of 20 unit cells. The enlarged view of the whole PnC region (**e**) and its unit cell (**f**) are also shown. Fabricated structure has the following parameters, measured using SEM : $w$ = 96 nm, $h$ = 90 nm, $a$ = 130 nm, $t$ = 22 nm, $r$ = 17 nm, $d$ = 70 nm.

We use free-standing 1D "block−tether" type PnCs supported by tapered waveguides, as schematically shown in Fig. 1a. This simple PnC structure is known to support complete bandgaps[34]. The unit cell of the PnCs (inset of Fig. 1a) is composed of an elliptical block and small tethers with thickness $d = 70$ nm, and lattice constant $a = 130$ nm. The major ($w$) and minor ($h$) axes of the elliptical block are 96 nm and 90 nm, respectively. To obtain a larger bandgap, we decrease the tether width ($t$) as much as possible (see Supplementary Fig. 3), leading to increased mass contrast between the elliptical block and the tether. We choose devices with a very small $t$ of ~20 nm for the experiments with SiV centers.

Figure 1b shows the dispersion relation (frequency vs. wave vector) for the phonon modes supported by the structure and obtained by simulating the unit cell with its experimentally measured dimensions (see Supplementary Section 2) using finite element method (COMSOL multiphysics). The band diagram features a large complete phononic bandgap with a width of 17.3 GHz centered at 59.1 GHz. These values are 4-10 times greater than those reported previously[19,20,35,36]. Similar to photonic bandgaps, the phononic bandgap results in a complete depletion of the phononic LDOS in this frequency range, as can be seen in the right panel. The robustness of the bandgap against fabrication imperfections is discussed in Supplementary Section 3. The bandgap frequency is chosen to overlap with the orbital ground state splitting ($\Delta_{GS}$) of a typical low-strain SiV center (>46GHz) and thus suppress the spontaneous single-phonon transitions, as detailed in the following section.

The suspended PnCs are fabricated into a single-crystal diamond substrate using top-down reactive ion etching followed by a quasi-isotropic etching step[37,38](see Methods and Supplementary Section 1). The single SiV centers are deterministically incorporated into the center of the array of PnCs by a mask implantation technique[39]. The scanning electron microscope (SEM) images of fabricated PnCs are shown in Fig. 1c-f. The PnC structure has a ~100 nm periodicity and features as small as ~20 nm (see Fig. 1f).

**Orbital lifetime measurements for SiV centers**

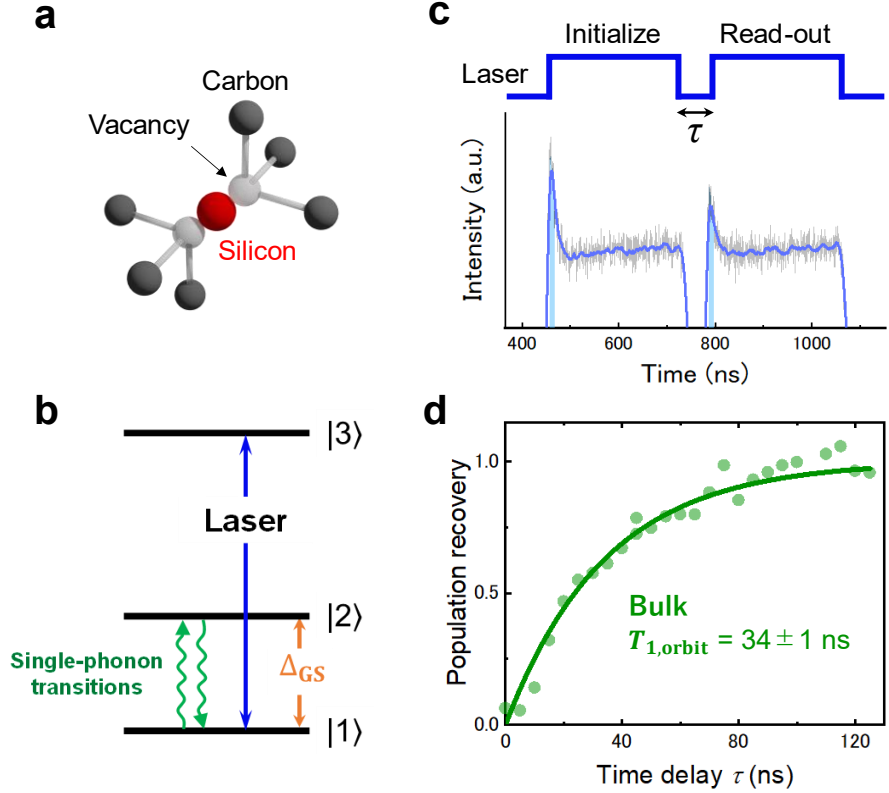

**Fig. 2 | Orbital lifetime measurements of single SiV centers. a,** Schematics of the atomic structure of the SiV center in diamond. **b,** Energy level structure of the SiV center, showing the frequency splitting between orbital branches in the ground state ($\Delta_{GS}$). The upward and downward spontaneous single-phonon processes between the ground states are illustrated as green arrows. The solid blue arrow indicates the laser resonant excitation to one of the energy levels for orbital lifetime measurements. **c**. Pulse sequence for the orbital lifetime measurements, consisting of 300ns-long initialization and read-out pulses with a time delay ($\tau$) in-between (upper panel). Example of a time-trace fluorescence spectrum measured at $\tau = 30$ ns (lower panel). The gray and blue lines are measured and smoothed data using a moving-average method to remove the noise. The blue areas indicate a 10 ns-wide integration range used to calculate the ratio between the leading edges and estimate population recovery. We note that the fluorescence increase in the steady state is due to power variations across the acousto-optic modulator carved pulse. **d**. Thermalization curve, showing the population recovery of state |1> (black dots), calculated as the ratio between the integrated leading edges of the read-out and initialization pulses in the time-trace spectrum shown in Fig. 2c as a function of $\tau$. The solid black line is a fitting curve with an exponential function to extract $T_{1,\text{orbit}}$ as the 1/e value. The extracted $T_{1,\text{orbit}}$ in bulk is $34 \pm 1$ ns.

The single SiV centers used to probe the phononic LDOS consist of a silicon atom and two vacancies occupying adjacent lattice sites in the diamond crystal, as schematically shown in Fig. 2a. Figure 2b displays the simplified electronic energy levels of a SiV center, with one of the excited states and two ground states with a splitting of $\Delta_{GS}$ (see Supplementary Section 6 for the detailed energy level). This splitting originates from the spin-orbit coupling in SiV centers and typical $\Delta_{GS}$ of unstrained SiV centers in bulk is ~50 GHz[40]. The transitions of single-phonon emission and absorption between the two ground states (green arrows) determine the lifetime of the orbital states ($T_{1,orbit}$), which also mainly limits the electronic spin coherence time of SiV centers around 4K[12]. To measure $T_{1,orbit}$ for our single SiV centers, we perform time-resolved pump-probe measurements (see Methods and Supplementary Fig. 5) with laser pulses resonant with one of the electronic transitions and probing population in state |1⟩ (blue arrow shown in Fig. 2b) at 4.4K. Figure 2c shows the pulse sequence where we measure the initialization and read-out of the orbital state separated by a variable delay $\tau$ through time-resolved fluorescence resulting from optical pumping (upper panel). The lower panel shows an example of a measured time-resolved spectrum for $\tau = 30$ ns. The ratio of the area under the leading edge of the read-out pulse to that of the initialization pulse (blue areas) measures the amount of population recovery from the initialized state (|2⟩) to the other ground state (|1⟩) via single-phonon transitions (green arrows). Figure 2d shows an example of population recovery as a function of $\tau$ measured for a single SiV center in bulk. From the fit of the exponential recovery (see Supplementary Section 7), we extract $T_{1,orbit}$ in bulk to be $34 \pm 1$ ns, which is comparable with that reported for bulk SiV centers around 4K[40].

**SiV orbital degrees of freedom as probes of phononic LDOS**

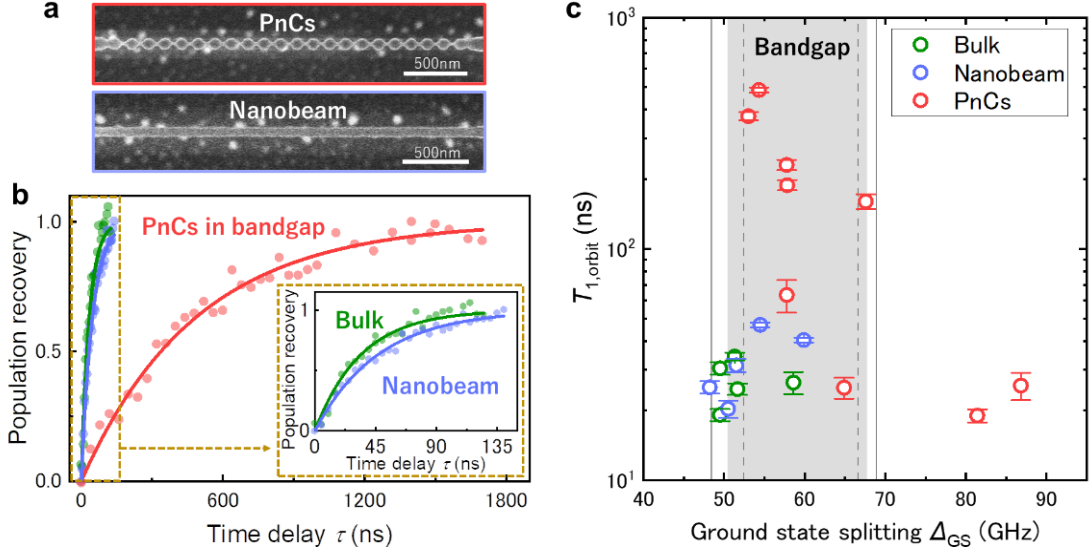

**Fig. 3 | Ground state splitting dependence of SiV orbital lifetimes. a,** SEM images of suspended diamond PnC (upper) and diamond nanobeam without PnC (lower) fabricated on the same chip. **b,** Thermalization curves measured for a single SiV center in bulk, nanobeam and PnCs. The solid lines are fitting curves based on a single exponential function. Inset shows the enlarged view of the thermalization curves for bulk and nanobeam. **c,** Measured $T_{1,\text{orbit}}$ as a function of $\Delta_{\text{GS}}$. The error bars represent the standard deviation of the fitting. The gray shaded region corresponds to the calculated phononic bandgap shown in Fig. 1e. The gray solid (dashed) lines indicate the uncertainty of the position and size of the bandgap induced by the standard deviation of -3nm (+3nm) from the measured tether width $t=22$nm.

Next, we compare the $T_{1,\text{orbit}}$ measured in PnCs to that in bulk as well as suspended diamond nanobeams with the same width and thickness as PnCs (Fig. 3a) and fabricated on the same chip. In this experiment, we use low-strain SiV centers with $\Delta_{\text{GS}} \sim 60$ GHz corresponding to our PnC bandgap. Using photoluminescence-based measurements (see Methods and Supplementary Section 8), we confirm that there is no significant spectral diffusion or linewidth broadening of SiV centers in PnCs. Figure 3b shows the thermalization curves measured for SiV centers of similar $\Delta_{\text{GS}}$ within the bandgap in a PnC (red), in a nanobeam (blue), and in bulk (green). We can clearly see that the population recovery observed for the SiV in PnCs is much slower than in bulk and in a nanobeam. The longest $T_{1,\text{orbit}}$ among the measured SiV centers within the bandgap is 486 ± 12 ns, which is approximately 18 (15) times longer than the average lifetime for 5 different SiV centers measured in bulk (nanobeam). This is clear evidence of the decrease of the phononic LDOS at ~ 60 GHz by the bandgap effect. On the other hand, as seen in Fig. 3b, we do not observe significant change in $T_{1,\text{orbit}}$ for the SiV centers in nanobeams relative to bulk despite the modified LDOS induced by their lower dimensionality[41]. This suggests that a complete bandgap is necessary to sufficiently suppress thermal phonons and the resulting phonon-induced relaxation.

We then use SiV centers with different $\Delta_{GS}$ to probe the extent of the phononic bandgap. Figure 3c shows the summary of $T_{1,orbit}$, overlaid with the same calculated phononic bandgap as the one shown in Fig. 1b, including the fabrication uncertainty of PnC parameters (see Supplementary Section 3). Overall, the values of $T_{1,orbit}$ for SiV centers with $\Delta_{GS}$ within the bandgap tend to be much higher than those in bulk and nanobeams, further confirming that the phonon-induced transitions between ground state orbitals are largely suppressed by the bandgap effect. The fluctuation of $T_{1,orbit}$ within the bandgap, particularly the lower $T_{1,orbit}$ values could be due to the variations in bandgap size and/or position caused by fabrication imperfection (see Supplementary Section 3) or the degradation of the emitters located near etched sidewalls[42]. We also measured $T_{1,orbit}$ on SiV centers with $\Delta_{GS}$ outside the bandgap (> 70GHz). As expected, $T_{1,orbit}$ outside the bandgap show similar values to those measured in bulk and nanobeams, confirming the realization of the large phononic bandgap of approximately 20 GHz around 60 GHz.

**Temperature-dependent measurements of orbital lifetimes**

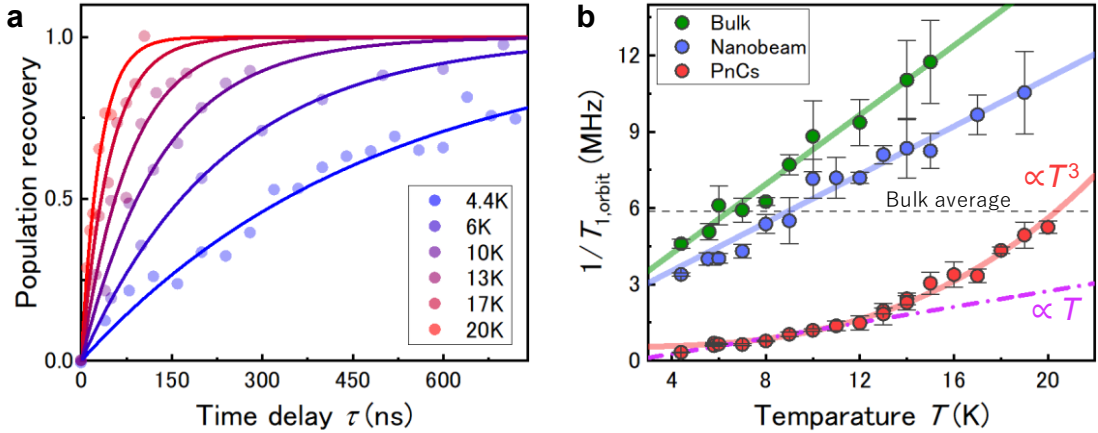

**Fig. 4 | Temperature dependence of orbital relaxation rates of single SiV centers. a,** Population recovery measured for a single SiV center in the PnC bandgap at 6 different temperatures. The solid lines are fitting curves based on a single exponential function to extract $T_{1,orbit}$. **b,** Temperature dependence of the measured orbital relaxation rate ($\Gamma=1/T_{1,orbit}$) for single SiV centers in bulk (black), nanobeam (green), and PnCs (red). In each case, the SiV center with the longest $T_{1,orbit}$ is used for temperature-dependent measurements. The error bars are standard deviations extracted from the single exponential fitting. The data for the bulk and the nanobeam can be fitted with a linear regression of $T$. For the SiV center in PnC, the dashed pink line is a linear fitting to the experimental data below 13K. The red solid curve is a fitting with a polynomial function with $T^3$. The dashed gray line indicates the average value of bulk SiV centers measured at 4.4K.

We finally investigate the temperature dependence of the population recovery by measuring $T_{1,\text{orbit}}$ from 4.4 K to 20 K. We use the same single SiV centers in bulk, nanobeam, and PnC as those shown in Fig. 3b. Figure 4a shows examples of the thermalization curves measured for the SiV center in the PnC bandgap at 6 different temperatures. As expected, the population recovery becomes faster with increasing sample temperature due to the higher thermal population of the residual phononic LDOS. The temperature dependence of the orbital relaxation rates ($\Gamma=1/T_{1,\text{orbit}}$) measured for SiV centers in bulk, nanobeam and PnC are plotted in Fig. 4b. The $\Gamma$ measured in bulk tends to increase linearly with temperature since single-phonon processes between the ground state levels are dominant below 20K, which is consistent with previously reported observations for bulk SiV centers[12,40]. We also clearly observe a linear temperature dependence of $\Gamma$ for the SiV center in a nanobeam, while showing a slightly slower speed of increase (0.47±0.02 MHz/K) compared to bulk (0.68 ± 0.04 MHz/K). The evolution of $\Gamma$ for the SiV center in a PnC below 13 K can also be approximated as a linear relationship with temperature, though it shows a much smaller slope of 0.15 ± 0.004 MHz/K. The clear difference in the rate of increase in $\Gamma$ between PnCs and bulk (nanobeam) originates from the differences in the phononic LDOS (see Supplementary Fig. S4), showing that modifying the LDOS is advantageous for suppressing the increase in $\Gamma$ at elevated temperatures. At temperatures above 12 K, $\Gamma$ for PnCs starts to show a clear deviation from the linear trend. Above this temperature, the experimental data can be well fitted with a function proportional to $T^3$ (red solid curve) as shown in Fig. 4b. This result indicates that higher-order phonon processes such as inelastic Raman processes[40] with frequencies outside the bandgap become dominant above 12K (see detailed discussion in Supplementary Section 9) as a result of the large suppression of the single-phonon processes by the PnC bandgap. These results show the successful demonstration of suppression of the spontaneous single-phonon transitions with high frequencies around 60 GHz by the PnC bandgap effect. Importantly, the measured $T_{1,\text{orbit}}$ for PnCs at 20K is ~ 30 ns, which is comparable to the average $T_{1,\text{orbit}}$ of bulk SiV centers (~ 27 ns) measured at 4.4K (a dashed gray line in Fig. 4b). This indicates that the PnCs with a complete bandgap can be useful to increase the operating temperature of SiV centers.

**Discussion**

In summary, we have demonstrated the suppression of high-frequency thermal phonons using PnCs and the mitigation of their detrimental effect on a target single quantum emitter in diamond. We fabricated free-standing diamond PnCs with sub-100 nm scale features to support a ~20 GHz-wide complete phononic bandgap centered at ~ 60 GHz. Using the PnCs, we showed that orbital relaxation rates of single SiV centers in diamond can be reduced by more than one order of magnitude compared to bulk and free-standing nanobeams. This large reduction in the relaxation rates due to inhibition of the single-phonon processes in the SiV center enables significant suppression of phonon-emitter interactions up to 20K. These results open the possibility to not only efficiently control specific high-frequency phonons of interest for applications in radio-frequency signal processing[43], optomechanics[44], and thermoelectrics[45], but also to improve optical and/or spin properties of solid-state quantum systems at higher temperatures, such as the spin coherence time of the SiV at 4K. Our approach to suppress the high-frequency thermal phonons and related phonon-induced relaxation could also be applied to any other solid-state quantum systems that interact with thermal phonons[24,46]. Finally, since our PnC structures are compatible with existing planar photonic and phononic integrated platforms[47], they could be a fundamental building block to realize phonon-mediated devices with solid-state quantum emitters and qubits for scalable quantum networks[48].

## Methods

### Device fabrication

We use a high-purity (nitrogen concentration less than 5 ppb) electronic grade single-crystal diamond (Element Six Corporation) that is etched using argon/chlorine followed by oxygen plasma to remove the surface-damaged layer caused by polishing. It is then cleaned using a 1:1:1 boiling tri-acid mixture of perchloric, nitric, and sulfuric acid. Silicon ions ($Si^+$) are implanted with an energy of 50 keV. The resulting mean ion range is ~35nm from the surface with a straggling of 8nm, as simulated by the software package "Stopping and Range of Ions in Matter" (SRIM). The SiV centers are then generated by a high-temperature (1100°C), high-vacuum annealing procedure followed by the same boiling tri-acid cleaning described above. The phononic crystals (PnCs) are patterned into the bulk diamond containing SiV centers by electron beam lithography, and dry etching. To realize free-standing PnCs, we use quasi-isotropic etching technique for undercutting the PhC slab. Details of the fabrication steps can be found in Supplementary Section 1.

### Photoluminescence measurements

The diamond sample is mounted on a closed-cycle liquid helium cryostat (attoDRY800) and cooled down to 4.4 K. A continuous-wave (CW) 520 nm laser diode (Thorlabs LP520-SF15) is used to pump the sample off-resonantly. The zero-phonon-line (ZPL) emission from single SiV centers in the sample is collected by an objective lens (0.9 NA ×100) and sent to a spectrometer equipped with a Si CCD camera to spectrally resolve the ZPLs. For photoluminescence excitation (PLE) measurements, we use a tunable CW Ti:sapphire laser (M-Squared SolsTis) for the resonant excitation to the SiV ZPLs. The PLE emission is collected from the phonon-sideband above 750 nm and is sent to an avalanche photodiode (APD) (Excelitas) to measure the photon counts. The frequency of the tunable laser is stabilized by continuous feedback with a wavemeter (High Finesse WS7). The CW 520nm laser is periodically pulsed by a digital delay generator (SRS DG645) to repump the SiV centers into the negatively charged state. A schematic illustration of the optical measurement setup can be found in Supplementary Section 5.

### Orbital lifetime measurements

We implement the two-pulse sequence to measure the orbital lifetime by pulsing our tunable CW Ti:sapphire laser with an acousto-optic modulator (AOM) (AA opto-electronic). The AOM is driven by an arbitrary waveform generator (Tektronix AWG70001A) through an RF switch (Mini-Circuits ZASWA2-50DR-FA+) with a voltage-controlled oscillator (Mini-Circuits ZX95-209-S+). The laser frequency is stabilized by continuous feedback from a wavemeter (High Finesse WS7). A pulsed 520 nm laser is also used to keep the negatively charged state of SiV centers. A time-correlated single-photon counting system (PicoHarp 300) with a superconducting nanowire single photon detector or APD is used to detect the photon counts from phonon-sideband emission of SiV centers. For temperature dependent measurements, the sample temperature is tuned using a local heater attached to the sample holder and it is then stabilized by continuous feedback with a proportional-integral-differential controller.

**Acknowledgments**
The authors would like to thank Prof. S. Iwamoto for their technical support and Prof. R. Anufriev, and Dr. R. Yanagisawa for their helpful discussions. This work was supported by ONR (Grant No. N00014-20-1-2425), ARO MURI (Grant No. W911NF1810432), NSF STC (Grant No. DMR-1231319), NSF ERC (Grant No. EEC-1941583). K.K. acknowledges financial support from JSPS Overseas Research Fellowships (Project No. 202160592). B.P. acknowledges support through a Marie Skłodowska-Curie fellowship from the European Union's Horizon 2020 research and innovation programme under the Grant Agreement No. 840968 (COHESiV). C.C. acknowledges support from A*STAR National Science Scholarship. This work was performed in part at the Center for Nanoscale Systems (CNS), Harvard University.

**Author contributions**
M.L. conceived the project. K.K. and C.C. performed the numerical simulations with help from M.H. K.K. designed the device. B.P. prepared the diamond sample. K.K. fabricated the devices with help from C.C. and W.D.. K.K., B.P., M.H., G.J., and C.J. carried out the measurement with help from D.A., C.X., M.Y., and N.S.. K.K. analyzed the data. K.K. and B.P. wrote the manuscript with contributions from all authors. M.L. supervised the project.

**Competing interests**
The authors declare no competing interests.

# Supplemental information for: Engineering Phonon-Qubit Interactions using Phononic Crystals

Kazuhiro Kuruma[1, a)], Benjamin Pingault[1,2], Cleaven Chia[1], Michael Haas[1], Graham D Joe[1], Daniel Rimoli Assumpcao[1], Sophie Weiyi Ding[1], Chang Jin[1], C. J. Xin[1], Matthew Yeh[1], Neil Sinclair[1], and Marko Lončar[1, b)]

[1] *John A. Paulson School of Engineering and Applied Sciences, Harvard University, Cambridge, Massachusetts 02138, USA*
[2] *QuTech, Delft University of Technology, PO Box 5046, 2600 GA Delft, The Netherlands, EU*

## 1. Phononic crystal fabrication

Before spin-coating of a 400nm-thick electron beam (EB) resist (ZEP520A), a 100 nm-thick SiN layer is deposited on the bulk diamond by plasma-enhanced chemical vapor deposition [Fig. S1a]. The SiN layer is used as a hard mask for etching the diamond later. The phononic crystal (PnC) pattern is written by EB lithography on the EB resist [Fig. S1b]. After the resist development [Fig. S1c], the SiN mask layer is etched by induced coupled plasma-reactive ion etching (ICP-RIE) with sulfur hexafluoride ($SF_6$) and octafluorocyclobutane ($C_4F_8$) gasses [Fig. S1d]. After removing the EB resist [Fig. S1e], we transfer the PnC pattern onto the diamond substrate by oxygen plasma RIE [Fig. S1f]. An $Al_2O_3$ layer with a thickness of 20nm is deposited by atomic layer deposition (ALD) for conformal coverage of the sample [Fig. S1g] and then etched out by following RIE with Ar and Cl gasses [Fig. S1h], while keeping the sidewalls of the PnCs covered with $Al_2O_3$. To realize free-standing structures, we employ a quasi-isotropic undercut technique[1] using an oxygen-based RIE [Fig. S1i]. The $Al_2O_3$ and SiN layers are finally removed using hydrofluoric acid (HF) [Fig. S1j].

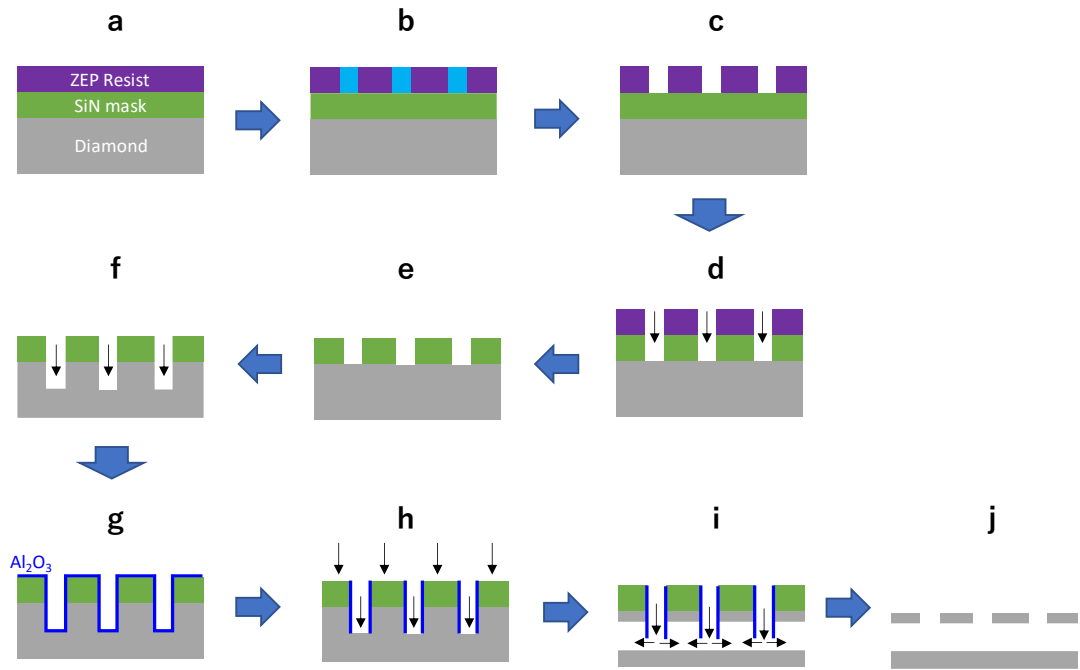

**Fig. S1 | Process flow for the phononic crystal fabrication. a**, A 100 nm-thick SiN is deposited on a diamond substrate by PECVD before spin coating of EB resist (ZEP520A). **b**, Phononic crystals are patterned into the resist by EB lithography. **c**, The EB resist is developed. **d**, The SiN hard mask is etched by ICP-RIE using the resist mask. **e**, The EB resist is removed. **f**, The pattern of the PnC is transferred into the diamond by oxygen plasma etching using the SiN mask. **g**, The $Al_2O_3$ layer is deposited on diamond by ALD. **h**, Top-down RIE using argon/chlorine gases is used to form the $Al_2O_3$ layer only on the sidewall of the diamond slab. **i**, Quasi-isotropic etching is performed based on oxygen plasma etching. **j**, The diamond sample is dipped into HF to remove $Al_2O_3$ and SiN.

## 2. Extraction of geometric parameters of fabricated PnCs

We fitted the PnCs in scanning electron microscopy (SEM) images to extract all geometric parameters: major and minor axes of the ellipse block ($w$ and $h$), tether width ($t$), radius of a circle at the rounded corner ($r$), and lattice constant ($a$), and thickness ($d$). As shown in the inset of Fig. S2, we fitted blocks (tethers) of the 73 unit cells with an ellipse (hyperbolic) function for $w$ and $h$ ($t$). 253 rounded corners of the 73 unit cells are fitted using circles for extracting $r$. We also extract $a$ from the distance between the centers of the fitted ellipses (53 samples). Figure S2 shows the histogram of the extracted $h$ as an example. The PhC thickness $d$ is also measured from 15 PnC devices in angled-view SEM images. Table S1 shows the summary of average and standard deviation (S.D.) values of extracted geometric parameters based on SEM observation.

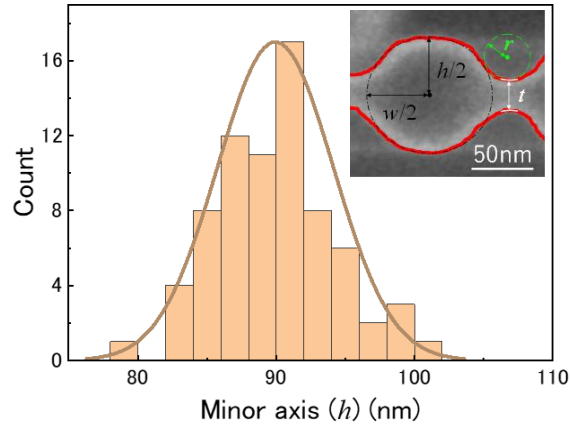

**Fig. S2 | Statistical evaluation of a geometric parameter.** Histogram of the extracted minor axes ($h$) of 73 unit cells. The solid orange line is the fitting result with a Gaussian function. Inset shows a SEM image for a unit cell with fitting curves for $h$, $w$ (black dashed), $t$ (green dashed) and $r$ (white solid). The red dots correspond to data points. The black and green dots indicate the center of the ellipse and circle fitting.

**Table S1 | Measured geometric parameters based on SEM observation.**

| Parameter | Average (nm) | S.D. (nm) |
|---|---|---|
| $w$ | 95.7 | 4.9 |
| $h$ | 89.9 | 4.2 |
| $t$ | 22.1 | 3.0 |
| $r$ | 16.9 | 5.6 |
| $a$ | 129.6 | 2.6 |
| $d$ | 70.3 | 3.7 |

## 3. Influence of fabrication imperfection on the phononic bandgap

Given the measured parameters from SEM shown in Table S1, we calculated the size and center position of the phononic bandgap by finite element method (FEM) simulations. The details of the calculation of band structure for diamond PnCs can be found in our previous work[2]. Figures S3a~f show the summary of the simulations for each parameter. The gray shaded area shows the range of the standard deviation. For the bandgap position, fluctuations in the tether width ($t$) and lattice constant ($a$) have the biggest influence, while the resulting changes of the position are very small (< 2%). For the bandgap size, variations in $t$ largely change the size by 18%. This is because the mass contrast between the elliptical block and tether is sensitive to the bandgap size (see Fig.S3c). We concluded that variations in the $t$ have the largest impact on the bandgap among measured parameters. From these results, we plotted the bandgap considering the measured variations in $t$, in Fig. 3c in the main text.

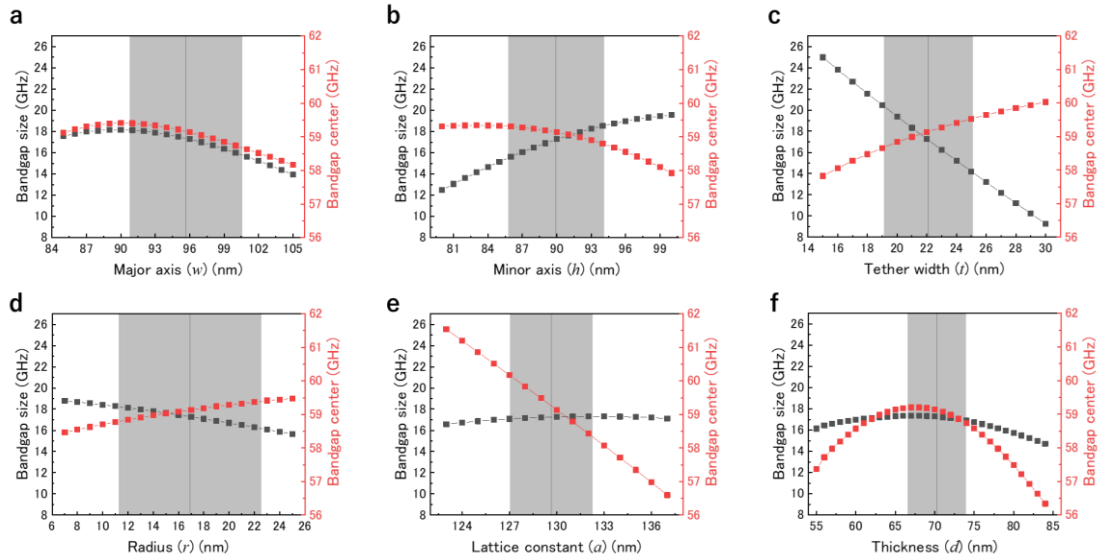

**Fig. S3 | Simulated bandgap size and center position for PnCs with measured variations. a**, $w$. **b**, $h$. **c**, $t$. **d**, $r$. **e**, $a$. and **f**, $d$. The solid black line indicates the average value. The gray shaded area corresponds to the standard deviation.

## 4. Calculation of phononic band structure and density of states for the nanobeam

We simulated the band diagram of the fabricated suspended nanobeam (beam width = 90 nm and thickness = 70nm), shown in Fig. 3a in the main text. Figure S4a shows the calculated band diagram of the nanobeam (orange solid line), overlaid with that of PnCs shown in Fig. 1b in the main text. Based on the simulated dispersion relations, we also calculated the corresponding phononic density of states (DOS)[3] for the nanobeam and PnCs, as shown in Fig. S4b. As can be seen in Fig. S4c, the DOS of the PnCs is completely depleted in the bandgap region centered at ~ 60GHz (gray shaded area in Fig. S4a), while that of the nanobeam has remained because there are two phononic modes over the bandgap region. This difference of the DOS results in the difference of $T_{1\text{orbit}}$ and its temperature dependence, as shown in Fig 3 and Fig 4 in the main text.

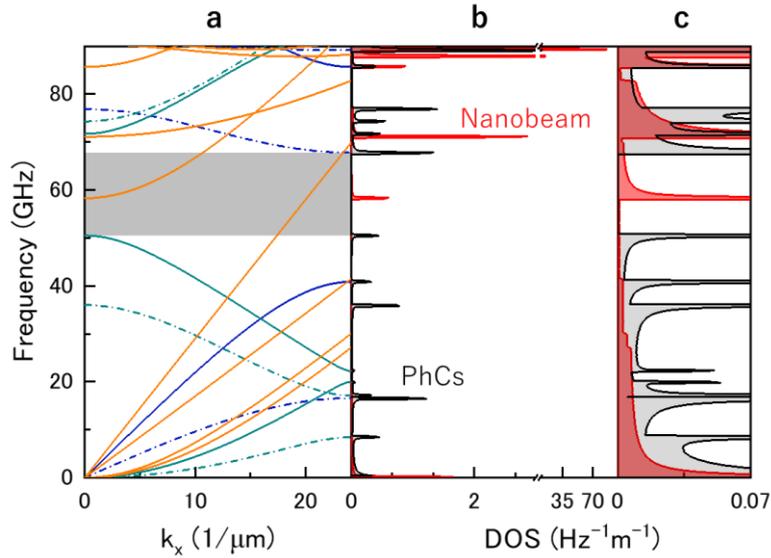

**Fig. S4 | Band structure and DOS for the nanobeam and PnCs. a**, Simulated band diagram for the nanobeam (orange solid curve), overlaid with that of PnCs shown in Fig. 1b in the main text. The gray shaded region corresponds to a complete bandgap of the PnCs. **b**, Corresponding calculated DOS for nanobeam (red) and PhCs (black). **c**, Enlarged view of **b** near the DOS value of zero.

## 5. Optical measurement setup

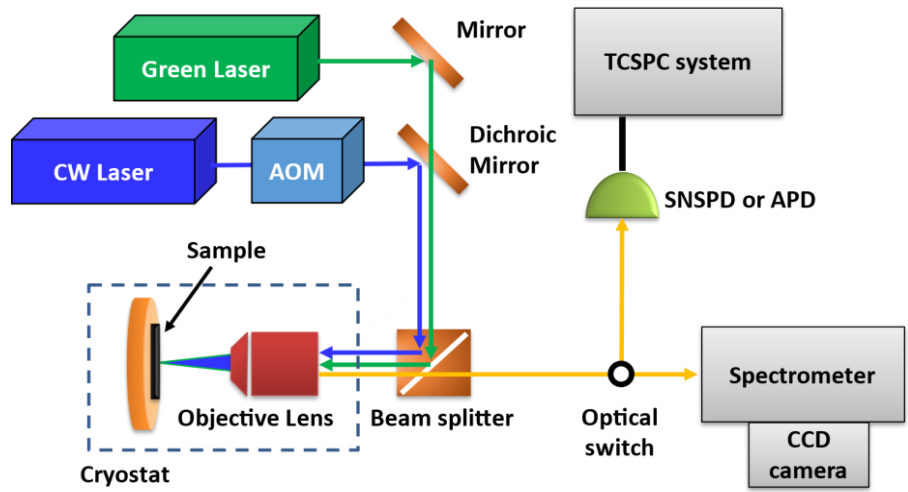

**Fig. S5 | Schematics of optical measurement setup.** AOM: acousto-optic modulator; SNSPD: superconducting nanowire single-photon detector; APD: avalanche photodiode; CCD: charged-coupled device; TCSPC: Time-correlated single photon counting.

## 6. Details of energy level structure of the SiV system

The detailed level structure of the SiV center under zero magnetic field is shown in Fig. S6a. The optical ground and excited state contain two orbital branches (black horizontal lines) with splittings of ΔES and ΔGS, respectively (blue arrows), which originates from the lifted degeneracy of the orbitals in each state by spin-orbit coupling. The orbital transitions in ground and excited state can be driven by the phonons with frequencies corresponding to the splittings (solid and dashed green arrows). This energy level configuration also provides four optical transitions in zero-phonon lines(ZPL) labeled as A, B, C, and D (red arrows) shows in Fig. S6a. Figure S6b shows an example of photoluminescence(PL) spectrum of a single SiV center measured in a PnCs by off-resonant excitation with a 520nm CW diode laser. For the extraction of $\Delta_{GS}$, we measure the distance between C and D lines observed in a PL or photoluminescence excitation (PLE) spectrum.

In order to confirm the single photon nature of our SiV centers, we perform second-order correlation measurements for a SiV center in one of the PnCs using a Hanbury Brown-Twiss setup equipped with two avalanche photodiodes (APDs). We resonantly excite its transition C by the tunable CW laser and collect the phonon-sideband fluorescence. Figure S7 shows the intensity correlation histogram measured at 4.3K. We fitted the data with a model[4] after convoluting it with a time resolution of our system (~550ps). The second-order correlation function at zero delay time, $g2(0)$, exhibits a clear antibunching with a value of 0.1, confirming that the SiV center is a single photon emitter.

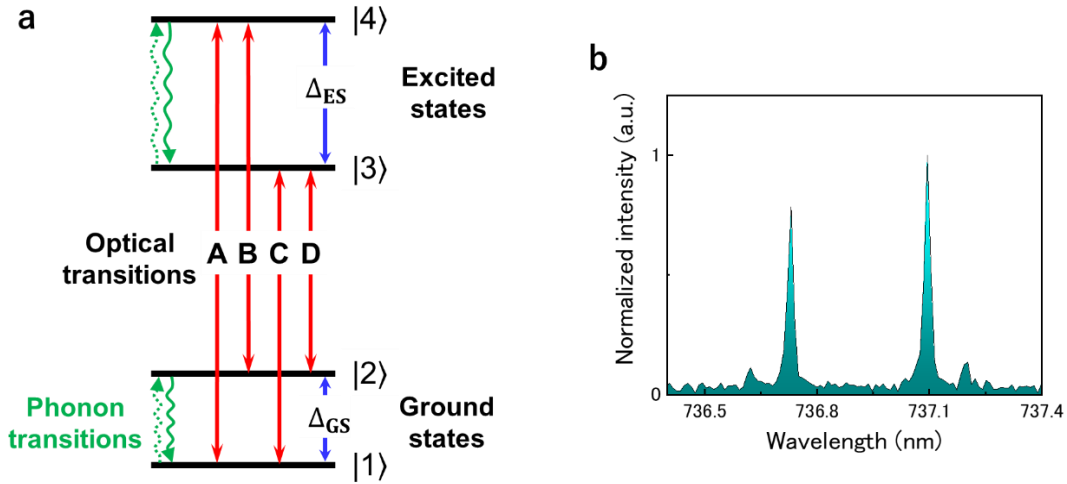

**Fig. S6 | Energy level structure of the SiV. a**, Schematics of electronic level structure of the SiV center with 4 zero-phonon transitions (labeled as A, B, C and D) under zero magnetic field. The ΔGS and ΔES indicate the ground and exticted state splittings, respectively. **b**, PL spectrum of a single SiV center in PnCs. The 4 sharp emission lines named as A, B, C and D corresponds to each optical transition shown in (a). ΔGS corresponds to the distance between C and D emission lines.

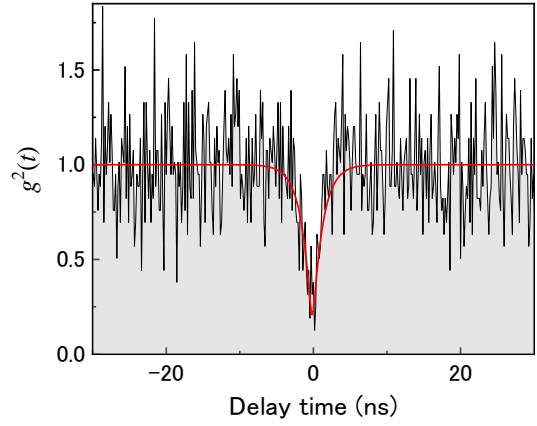

**Fig. S7 | Second-order correlation measurements**. Normalized auto-correlation function g²(t) measured for a SiV center in a PnC sample. The red solid line corresponds to a fitting curve considering the time response of our detection system.

## 7. Extraction of orbital lifetime $T_{1,\text{orbit}}$

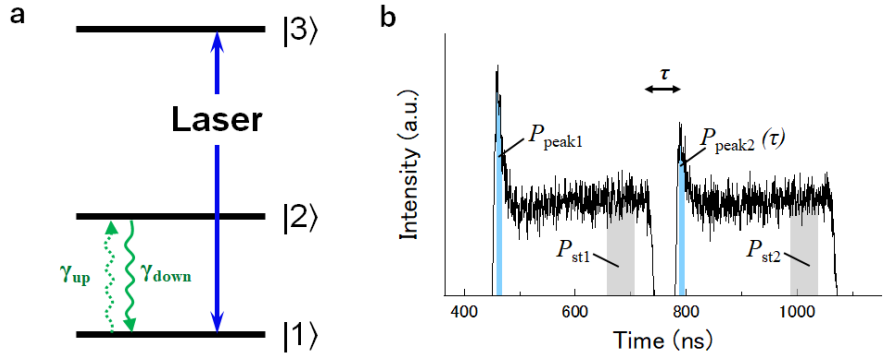

**Fig. S8 | Orbital lifetime measurements. a**, Simplified energy level structure of a SiV center. **b**, Time-resolved spectrum for time delay $\tau$ of 30ns.

In time-resolved pump-probe experiments to measure $T_{1,\text{orbit}}$, we used a pulsed laser resonant with one of the electronic transitions to probe the population in state |1> (blue arrow shown in Fig. S8a). Figure S8b shows an example of a time-resolved fluorescence spectrum measured with a time delay $\tau$ = 30 ns between two laser pulses. The blue area under the leading edge of the first (pump) pulse corresponds to the thermal population $P_{\text{peak1}}$ in one of the ground state levels (|2>). The population then decays to a stationary state, whose population $P_{\text{st1}}$ (light gray area) is determined by a competition between the optical pumping rate and orbital relaxation rates between ground states ($\gamma_{\text{up}}$ and $\gamma_{\text{down}}$). The blue area under the leading edge of the second (probe) pulse after a time interval of $\tau$ corresponds to partially recovered population $P_{\text{peak2}}(\tau)$ to the other ground state |1> due

to thermalization processes. The stationary state population for the second pulse is also defined as $P_{st2}$. From the peak ratio $(P_{peak2}(\tau) - P_{st2})/(P_{peak1} - P_{st1})$ as a function of time delays ($\tau$) shown in Fig. 2c in the main text, we can extract the orbital lifetime of $T_{1,\text{orbit}}$ (= $(\gamma_{up} + \gamma_{down})^{-1}$) using the following equation[5]:

$$Peak\ ratio = \frac{P_{peak2}(\tau) - P_{st2}}{P_{peak1} - P_{st1}} = 1 - exp\left(-\frac{\tau}{T_{1,orbit}}\right) \qquad (S1)$$

## 8. Saturation response and stability of a SiV ZPL transition in PnCs

We investigate the saturation response of a SiV ZPL transition in phononic crystals (PnCs). We choose the SiV with a ground state splitting of ~58GHz, exhibiting $T_{1,orbit}$ of 183 ns shown in Fig. 3c in the main text. We scan the C transition of the SiV using a continuous wave (CW) laser to measure its PLE spectrum (inset of Fig. S9a). A pulsed 520 nm laser is also used to periodically repump the SiV into the negatively charged state. Figure S9a shows a summary of the integrated counts of measured PLE spectra as a function of the laser power. The SiV transition saturates around 1$\mu$W.

We then check the stability of the same SiV embedded in PnCs by measuring its C transition over a ~2.5 hours period using the same CW laser. The laser power is set to be 0.3 $\mu$W, which is below the saturation power shown in Fig. S9a. The color map of PLE spectra is shown in Fig S9b. We do not observe any significant spectral diffusion and blinking of the SiV transition even in PnCs. Figure S9c shows the time-averaged PLE spectrum. By fitting with a Lorentzian function, the linewidth of the spectrum is deduced to be 387 MHz.

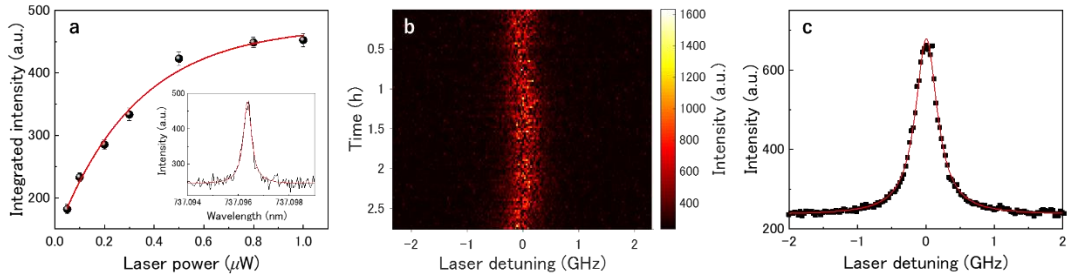

**Fig. S9 | Saturation curve and stability of a SiV in PnCs. a**, Saturation curve measured for a C transition of a SiV in PnCs using the CW laser. The red solid line is a fitting curve with an exponential function. Inset shows a time-averaged PLE spectrum of 10 times scans measured at 0.1 $\mu$W. Scale bar is one standard deviation. **b**, Color map of the C transition taken at 0.3 $\mu$W. **c**, Time-averaged PLE spectrum of 100 times scans. The single PLE scan takes about 90s.

## 9. Model of electron-phonon interactions

For describing the temperature dependence of orbital relaxation rates shown in Fig. 4b in the main text, we first consider the first-order electron-phonon transitions between orbital states in SiV centers, involving spontaneous absorption or emission of a single-phonon with the resonance to the frequency of the ground state splitting, as shown in Fig. S10a. The rates of upward $\gamma_+$ (absorption) and downward $\gamma_-$ (emission) phonon transitions are given by[6]

$$\gamma_+ = 2\pi\chi\rho\Delta^3 n(\Delta, T) \qquad (S2)$$
$$\gamma_- = 2\pi\chi\rho\Delta^3 [n(\Delta, T) + 1] \qquad (S3)$$

where, $\chi$ is a coupling constant that encapsulates averaged interaction over all phonon modes and polarizations, while $\rho$ is the phonon density of states. $n$ indicates the occupation of phonon modes following the Bose-Einstein distribution. In the temperature $T > h\Delta_{GS}/k_B$, the equations above (S2 and S3) can be approximated as follows:

$$\gamma_+ \approx \gamma_- \approx \frac{2\pi}{\hbar}\chi\rho\Delta^2 k_B T \qquad (S4)$$

Hence, the orbital relaxation rates are linearly proportional to temperatures $T$. Based on this model, we used a simple linear regression (= A + B$T$, where A and B are constant) to fit experimentally measured relaxation rates in bulk, nanobeam and PhC shown in Fig. 4b in the main text. The summary of fitting results is shown in Table S2.

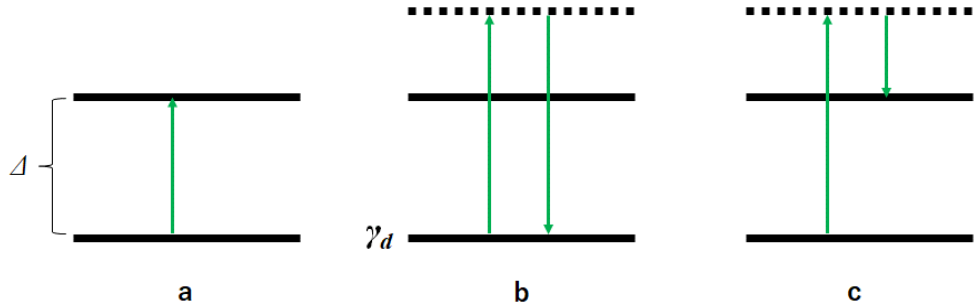

**Fig. S10 | Various phonon transitions between orbital states in SiV centers. a**, Single-phonon direct transition. Two-phonon transitions via elastic Raman process (**b**) and inelastic Raman process (**c**). $\Delta$ and $\gamma_d$ and indicate the splitting between the orbital states and the rate for a pure dephasing of the state, respectively.

**Table S2 | Fitting results using a linear regression (A + B$T$) for orbital relaxation rates in bulk, nanobeam and PnCs.**

|  | A (MHz) | B (MHz/K) |
|---|---|---|
| Bulk | 1.47±0.36 | 0.68±0.04 |
| Nanobeam | 1.62±0.34 | 0.47±0.02 |
| PnCs | -0.35±0.04 | 0.15±0.004 |

In Fig. 4b in the main text, the orbital relaxation rates of PnCs above 12 K exhibits clear deviations from the linear scaling with temperatures of $T$, suggesting that the dephasing of the orbitals states is determined by high-order phonon processes involving two-phonons rather than the single-phonon process owing to the large suppression of single phonon direct transitions by the PnC bandgap effect.

Similarly, a nonlinear broadening of their ZPL linewidth at temperatures above 20 K is reported in a previous work using SiV centers[6]. This observation is well described by the scaling of $T^3$, stemming from two-phonon process via elastic Raman process (Fig. S10b), which is more dominant than inelastic Raman process (Fig. S10c) in SiV centers. The orbital relaxation rate for elastic Raman process is expressed as follows[6]:

$$\gamma_{d-} \approx \gamma_{d+} \approx 2\pi\hbar^2\chi^2\rho^2\Delta^2 \int_0^\infty n(\omega,T)(n(\omega,T)+1)\omega^2 d\omega$$
$$= \frac{2\pi^3}{3\hbar}\chi^2\rho^2\Delta^2 k_B^3 T^3 \quad (S5)$$

where $\omega$ is frequency. The final form of the equation (S5) shows the orbital relaxation rates scales as $T^3$. We fitted the measured rates with a function of $T^3$ (=A+B$T^3$, where A and B are constant), and obtained A and B to be 0.52 ± 0.06 MHz and 6.3×10$^{-4}$±1.8×10$^{-5}$ MHz K$^{-1}$, respectively. As can be seen in Fig. S11, the experimental data can be well described by the model with the scaling of $T^3$ dependence. On the other hand, other potential models[7,8] involving a two phonon Raman process with the scaling of $T^5$ or $T^7$ increase the fit error and cannot explain the experimental data well.

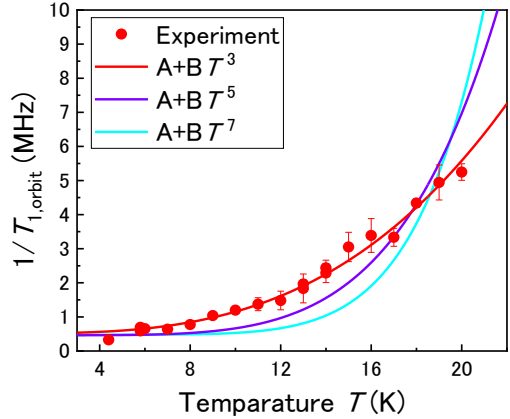

**Fig. S11 | Measured orbital relaxation rates of a SiV center in PnCs as a function of sample temperatures.** The red dots are experimental data. The red, purple, and light blue curves correspond to fitting results using a model with the scaling of $T^3$, $T^5$, $T^7$, respectively.